\documentclass[twocolumn, twocolappendix]{aastex631}

\usepackage{amsmath} 
\usepackage{multirow}
\usepackage{longtable}

\usepackage[normalem]{ulem}

\begin{document}

\title{SPARC4 control system}

\author[0000-0001-7148-6886]{Denis Bernardes}
\affiliation{Instituto Nacional de Pesquisas Espaciais, Avenida dos Astronautas, 1.758, Jardim Granja, São José dos Campos, São Paulo, Brazil}
\author[0000-0002-8620-8149]{Orlando Verducci Junior}
\affiliation{Laboratório Nacional de Astrofísica, Rua dos Estados Unidos, 154, Bairro das Nações, Itajubá, Minas Gerais, Brazil.}
\author[0009-0006-5519-3446]{Francisco Rodrigues}
\affiliation{Laboratório Nacional de Astrofísica, Rua dos Estados Unidos, 154, Bairro das Nações, Itajubá, Minas Gerais, Brazil.}

\author[0000-0002-9459-043X]{Claudia V. Rodrigues}
\affiliation{Instituto Nacional de Pesquisas Espaciais, Avenida dos Astronautas, 1.758, Jardim Granja, São José dos Campos, São Paulo, Brazil}
\author[0000-0003-0680-1979]{Luciano Fraga}
\affiliation{Laboratório Nacional de Astrofísica, Rua dos Estados Unidos, 154, Bairro das Nações, Itajubá, Minas Gerais, Brazil.}
\author[0000-0002-5084-168X]{Eder Martioli}
\affiliation{Laboratório Nacional de Astrofísica, Rua dos Estados Unidos, 154, Bairro das Nações, Itajubá, Minas Gerais, Brazil.}
\author{Clemens D. Gneiding}
\affiliation{Laboratório Nacional de Astrofísica, Rua dos Estados Unidos, 154, Bairro das Nações, Itajubá, Minas Gerais, Brazil.}
\author[0009-0004-2014-1852]{André Luiz de Moura Alves}
\affiliation{Laboratório Nacional de Astrofísica, Rua dos Estados Unidos, 154, Bairro das Nações, Itajubá, Minas Gerais, Brazil.}
\author{Juliano Romão}
\affiliation{Laboratório Nacional de Astrofísica, Rua dos Estados Unidos, 154, Bairro das Nações, Itajubá, Minas Gerais, Brazil.}
\author[0000-0001-9200-3441]{Laerte Andrade}
\affiliation{Laboratório Nacional de Astrofísica, Rua dos Estados Unidos, 154, Bairro das Nações, Itajubá, Minas Gerais, Brazil.}

\author[0000-0001-8179-1147]{Leandro de Almeida}
\affiliation{Laboratório Nacional de Astrofísica, Rua dos Estados Unidos, 154, Bairro das Nações, Itajubá, Minas Gerais, Brazil.}
\author[0000-0002-0337-1363]{Ana Carolina Mattiuci}
\affiliation{Instituto Nacional de Pesquisas Espaciais, Avenida dos Astronautas, 1.758, Jardim Granja, São José dos Campos, São Paulo, Brazil}

\author{Flavio Felipe Ribeiro}
\affiliation{Laboratório Nacional de Astrofísica, Rua dos Estados Unidos, 154, Bairro das Nações, Itajubá, Minas Gerais, Brazil.}
\author[0000-0002-7095-4147]{Wagner Schlindwein}
\affiliation{Instituto Nacional de Pesquisas Espaciais, Avenida dos Astronautas, 1.758, Jardim Granja, São José dos Campos, São Paulo, Brazil}
\author[0009-0002-6323-0172]{Jesulino Bispo dos Santos}	  		
\affiliation{Laboratório Nacional de Astrofísica, Rua dos Estados Unidos, 154, Bairro das Nações, Itajubá, Minas Gerais, Brazil.}
\author[0000-0002-0386-2306]{Francisco Jose Jablonski}
\affiliation{Instituto Nacional de Pesquisas Espaciais, Avenida dos Astronautas, 1.758, Jardim Granja, São José dos Campos, São Paulo, Brazil}
\author[0000-0002-2075-2424]{Julio Cesar Neves Campagnolo}
\affiliation{Centro Federal de Educação Tecnológica Celso Suckow da Fonseca, Rio de Janeiro, Rio de Janeiro, Brazil}
\author{Rene Laporte}
\affiliation{Instituto Nacional de Pesquisas Espaciais, Avenida dos Astronautas, 1.758, Jardim Granja, São José dos Campos, São Paulo, Brazil}
\begin{abstract}

SPARC4 is a new astronomical instrument developed entirely by Brazilian institutions, currently installed on the 1.6-m Perkin-Elmer telescope of the Pico dos Dias Observatory. It allows the user to perform photometric or polarimetric observations simultaneously in the four SDSS bands (g, r, i, and z). 
In this paper, we describe the control system developed for SPARC4. This system is composed of S4ACS, S4ICS, and S4GUI softwares and associated hardware. S4ACS is responsible for controlling the four EMCCD scientific cameras (one for each instrument band). S4ICS controls the sensors and motors responsible for the moving parts of SPARC4. Finally, S4GUI is the interface used to perform observations, which includes the choice of instrument configuration and image acquisition parameters. S4GUI communicates with the instrument subsystems and with some observatory facilities, needed during the observations. Bench tests were performed for the determination of the overheads added by SPARC4 control system in the acquisition of photometric and polarimetric series of images. In the photometric mode, SPARC4 allows the acquisition of a series of 1400~full-frame images, with a deadtime of 4.5~ms between images. Besides, several image series can be concatenated with a deadtime of 450~ms plus the readout time of the last image. For the polarimetric mode, measurements can be obtained with a deadtime of 1.41~s plus the image readout time between subsequent waveplate positions. For both photometric and polarimetric modes, the user can choose among operating modes with image readout times between 5.9~ms and 1.24~s, which ultimately defines the instrument temporal performance.
\end{abstract}


\section{Introduction} \label{sec:intro}

\textit{Instituto Nacional de Pesquisas Espaciais} (INPE) in collaboration with \textit{Laboratório Nacional de Astrofísica} (LNA) developed a new astronomical instrument named Simultaneous Polarimeter and Rapid Camera in Four Bands \citep[SPARC4 - ][]{Rodrigues2012, Claudia_SAB_2024}, which is currently installed on the 1.6-m Perkin-Elmer Telescope of \textit{Pico dos Dias} Observatory (OPD, in Portuguese). SPARC4 allows simultaneous photometric and polarimetric acquisitions in four spectral bands, which are similar to the g, r, i, and z bands of the Sloan Digital Sky Survey (SDSS) system \citep{SDSS}. The four scientific detectors are iXon Ultra 888 Electron Multiplying Charged Coupled Devices~(EMCCDS), produced by the Oxford Instruments company. These devices offer the Electron Multiplying feature for amplifying signals from faint objects. Furthermore, they support the Frame Transfer (FT) option, enabling acquisition rates of up to 27 frames per second (fps) for full-frame images (1024~$\times$~1024~pixels). Besides, these devices have a coating and a window optimized for the spectral band in which they operate. For polarimetric observations, SPARC4 uses a dual-beam technique, in which the incident light is split into ordinary and extraordinary components in the  polarimetric module and simultaneously acquired by the EMCCD cameras. 
The adopted polarimetric technique was already extensively used in another OPD instrument named \citep[IAGPOL, ][]{IAGPOL_1996}, which has a significant contribution to the scientific production of the observatory. The SPARC4 photometric system is very similar to the SDSS one: preliminary results are presented by \cite{Wagner_SAB_2024}. The SPARC4 characteristics are suitable for the study of variable objects such as cataclysmic variables, binaries systems, and exoplanets, to cite some scientific cases. The first light of the instrument occurred on November~4, 2022, and it was subsequently delivered to the observatory in semester 2024A. More information about SPARC4 can be found on its website\footnote{\url{https://coast.lna.br/home/sparc4}} and in other papers in preparation.

Different instrument designs have been developed over the years according to the scientific case they were intended to address. In the following, some examples of instruments considered relevant for this paper are presented. These instruments share at least one characteristic with SPARC4. This comparison is essential to position SPARC4 in the broader context of its counterparts. 
The Double Image Polarimeter - Ultra Fast \citep[DIPol-UF, ][]{Piirola2021} is a polarimeter installed on the 2.5~m Nordic Optical Telescope (NOT), at the Roque de Los Muchachos Observatory, Canary Island. DIPol-UF allows for the acquisition of simultaneous polarimetric measurements for the BVR bands. For that, three iXon Ultra 897 EMCCD cameras made by Oxford Instruments, with 512~$\times$~512~pixels of 16~$\mu$m were used. Using DIPOL-UF, a polarimetric precision of 10$^{-5}$ or better could be achieved, for stars with magnitudes in the V band ranging from 3.77 to 7.65, for a total exposure time of 30 to 40~minutes. Besides, Dipol-UF provides a maximum acquisition rate for linear polarimetry of 1.5~Hz.
The Parallel Imager for Southern Cosmology Observations \citep[PISCO, ][]{Stalder2014} is a photometer installed on the 6.5~m Magellan telescope, at the Las Campanas Observatory, designed for fast and synchronous acquisitions of low-brightness objects, in the g, r, i, and z bands of the SDSS system. For data acquisition in each band, a MITLL CCID34 CCD camera, with a resolution of 3072~x~6144~pixels of 10~$\mu$m is used, providing a FoV of 9~arcmin². PISCO has no polarimetric capabilities. 
the Multicolour OPTimised Optical Polarimeter (MOPTOP) instrument \citep{Jermak2018} is a polarimeter currently installed on Liverpool Telescope, which uses the same dual-beam technique of SPARC4 for splitting the light into the ordinary and extraordinary rays. Each of these rays are acquired by an Andor Zyla 4.2 sCMOS detector with a sensor of 4.2 megapixels of 6.5~$\mu$m, providing a FoV of 7~arcmin². Using MOPTOP, \cite{Bernardes_2024} determined an upper limit for the GRB 230818A polarization.

In this paper, we present the control system developed for the SPARC4 instrument. This document is organized as follows. Section~\ref{sec:instrument} presents an overview of the entire hardware of the instrument. Section~\ref{sec:acq_system} describes the control system and presents the procedure adopted for the determination of the overheads added by the instrument during the acquisition of photometric and polarimetric series of images. Finally, Section~\ref{sec:conclusion} presents our conclusions.


\section{Hardware description} \label{sec:instrument}

\subsection{Optical components} \label{subsec:optical_comp}

Following the optical path, the light exits the telescope, passes through the autoguider module, and reaches the polarimetric module. This polarimetric module is composed of the polarimetric calibration wheel, the retarder waveplates, and the analyzer. In the calibration wheel, there are five slots, one allocated for a polarizer and another for a depolarizer, which are used to measure instrumental polarization and polarization efficiency. Regarding the retarder waveplates, SPARC4 allows the user to choose between a half-wave plate, for linear polarization measurements, or a quarter-wave plate for circular and linear polarization measurements. The analyzer is a Savart prism. The polarimetric module allows for the removal of any polarimetric optics from the optical path to minimize photon loss in photometric acquisitions. Then, the light passes through the optical module, where it reaches a collimator and is split into four beams (channels) by three dichroic mirrors. A fourth mirror is used to achieve a symmetric mechanical distribution of the detectors. Each collimated beam reaches a focuser that transforms it into a f/5 beam. Finally, each beam encounters its respective detector, which are described in Section \ref{subsec:detectors}.

\subsection{Mechanisms} \label{subsec:mechanisms}

\begin{figure*}[tbh]
    \centering
    \includegraphics[width=\textwidth]{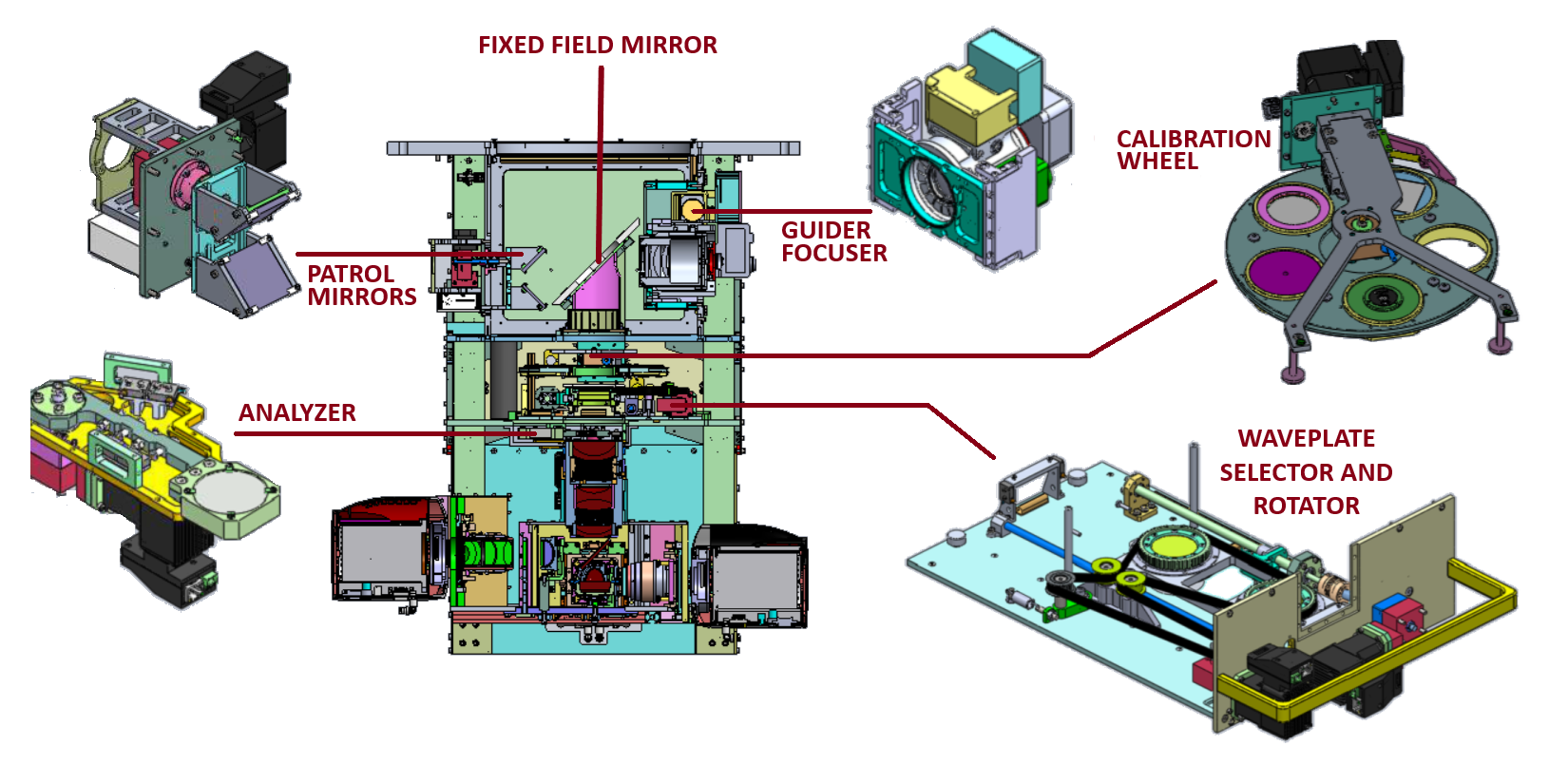}
    \caption{SPARC4 mechanical design showing the mechanisms of the instrument.}
    \label{fig:Mechanisms}
\end{figure*}

Figure \ref{fig:Mechanisms} shows the SPARC4 mechanical design, emphasizing the mechanisms. These mechanisms control the patrol mirrors of the autoguider, the autoguider focuser, the calibration wheel, the waveplate selector, the waveplate rotator, and the analyzer. The patrol mirror system is used to select a sky region to perform the autoguiding and to direct the light to the autoguider detector, an Andor Luca camera. This is accomplished by two rotating mirrors that reflect a section of a fixed-field mirror towards the camera. This fixed mirror has a central hole for the passage of the central telescope beam, so it reflects to the autoguider camera an annular region around the scientific FoV. The guiding camera is controlled by the AutoGuider System (AGS), a software developed by the observatory to assist the telescope in tracking a star during an observation. The focuser is a linear mechanism for the adjustment of the guiding camera focus. This is necessary as the telescope focus changes for each SPARC4 operation mode (see Section \ref{operation modes}). The patrol mirror together with the focuser make up the guiding system of SPARC4. 

The calibration wheel has five positions: clear, polarizer, depolarizer, a dark plate to serve as a shutter, and another position available for a possible new element. The waveplate selector provides three operational positions: clear, half-wave plate, and quarter-wave plate. The waveplate rotator is responsible for the control of the waveplate position angle. This angle can be set to sixteen different predefined positions, varying from 0\degr\ up to 337.5\degr\, with steps of 22.5\degr. Finally, the analyzer mechanism is a rotating arm with a 30\degr\ stroke to place the Savart prism on the optical path. The set of calibration wheel, waveplate selector, waveplate rotator, and analyzer mechanisms makes up the polarimetric module of SPARC4.

\subsection{Scientific detectors} \label{subsec:detectors}

SPARC4 uses four FT EMCCD iXon Ultra 888 cameras, produced by Oxford Instruments. These devices use CCD201-20 sensors of 1024~$\times$~1024~pixels of 13~$\mu$m, produced by Teledyne E2V company resulting in a FoV of 5.6~arcmin$^2$. With a maximum readout rate of 30~MHz, these cameras are capable of providing an acquisition rate of up to 27 fps for full-frame images. Also, using the EM option, an amplification gain of up to 300~e-/e- can be achieved, allowing the observation of relatively faint targets. In the slower conventional (not EM) readout mode, these devices provide a read noise of 3.5~e-/pix, for a temperature of -80~\degr C. Furthermore, the windows and coatings of each SPARC4 CCD camera were optimized for their spectral band operation and to minimize the fringing effect in red and infrared. See \cite{Bernardes_2018} for a detailed characterization of the SPARC4 cameras.

\subsection{Computing system}

The SPARC4 computing system is located in the OPD data center on the ground floor of the 1.6-m Perkin-Elmer telescope building, approximately 15 meters from the telescope located on the third floor. This data center is climate controlled and equipped with an uninterruptible power supply (UPS) to ensure stable operations.

The system comprises five rack-mounted servers and a desktop computer. One server is designated for hosting the user interface software and controlling the instrument moving mechanisms, while the remaining four servers are dedicated to running the image acquisition software S4ACS that interfaces with each detector via a USB 3.0 optical fiber extender. This configuration guarantees reliable high-speed data transmission without signal degradation, accommodating the long-distance connections necessary for effective operation. Detailed descriptions of the software applications will be provided in subsequent sections. Additionally, the desktop computer is utilized for real-time data analysis during observations using IRAF/PyRAF.

The SPARC4 network infrastructure includes two gigabit routers, each serving a distinct local network: one dedicated to communication and the other exclusively for data transfer. Each SPARC4 computer is equipped with a dual-port 1 gigabit Ethernet network interface card, with each port linked to one of the local networks to ensure reliable data and communication pathways. Furthermore, a switch located near the instrument connects to the communication router via an optical fiber link, effectively integrating it into the communication network. This switch also connects to individual motors and two remote power units via Ethernet cables, managing power for the detectors, USB 3.0 optical fiber extenders, and motors.

Each server operates on Windows Server and is equipped with an Intel processor, 16 GB of RAM, and a dedicated 480 GB SSD for the operating system. Additionally, each of the four acquisition servers features a 2~TB SSD for image data storage, providing efficient write performance required for high data acquisition rates, along with two 2~TB HDDs for backup purposes. The desktop computer runs Linux Fedora Workstation Edition and is equipped with an Intel Core i7 processor, 32 GB of RAM, a 4 TB NVMe SSD, and an additional 10~TB HDD for backup. The image storage disks from the acquisition servers are mounted as read-only on the desktop computer using the NFS protocol over a dedicated data-transfer network that is isolated to optimize data handling.

Each image acquisition server achieves precise temporal synchronization through USB-connected Meinberg devices, which receive IRIG time code signals from the observatory's time server. This timing system ensures submillisecond precision for timestamps embedded in the FITS headers of all acquired SPARC4 images. The observatory's time server is a rack server equipped with a Meinberg PCIe card, located in the OPD data center. The time reference is ultimately derived from a GPS antenna mounted on the roof of the telescope building. To connect the rooftop GPS antenna to the data center, the GPS signal is converted and transmitted via optical fiber using Meinberg GOAL modules, which maintain the integrity of the signal throughout the extended cable run.

\section{Control system} \label{sec:acq_system}

\begin{figure*}[tbh]
    \centering
    \includegraphics[width=\textwidth]{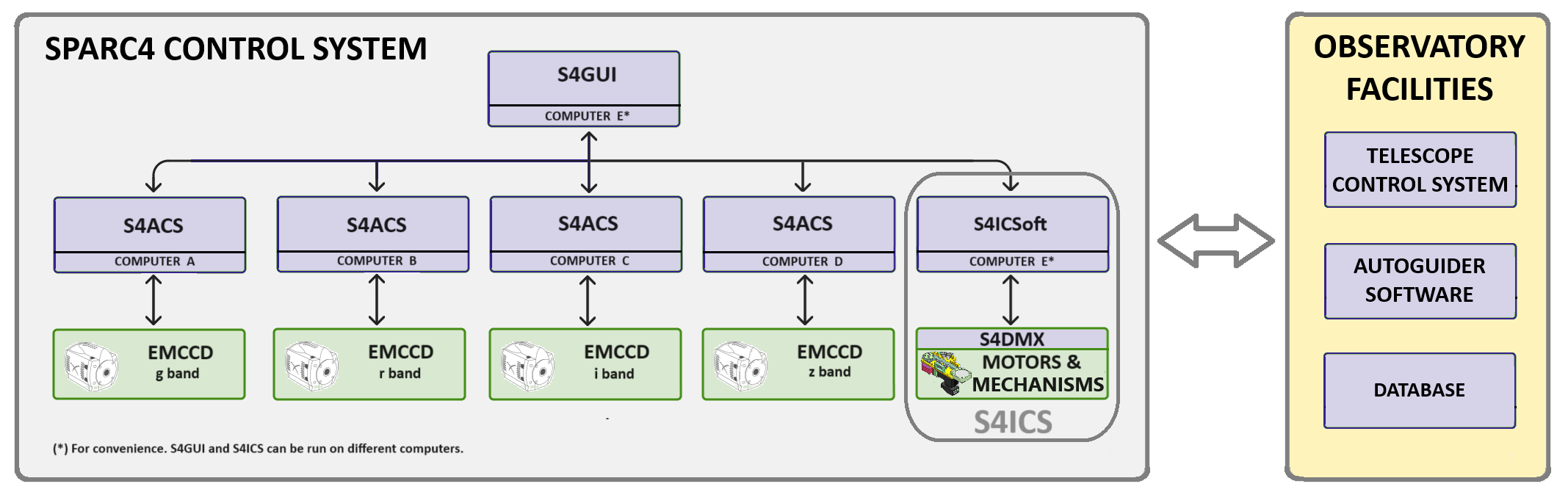}
    \caption{Diagram of the SPARC4 control system. In this diagram, the background colors distinguish the software (purple) from the hardware (green) components and the arrows represent the flux of the information. The names inside boxes are described in the text. The box at the right shows some of the other observatory softwares that interact with the SPARC4 system.}        
    \label{fig:sistema_controle_SPARC4}
\end{figure*}

SPARC4 is composed of a series of subsystems that need to work together to ensure the proper operation of the instrument. The control of each subsystem is done using a dedicated software. These softwares are the Acquisition Control System~(S4ACS), the Instrument Control Software (S4ICSoft), and the Graphical User Interface (S4GUI). S4ACS is responsible for controlling the scientific cameras. S4ICSoft, together with S4DMX, another software stored in motors, are responsible for controlling the mechanisms of SPARC4. S4GUI is a user interface that coordinates the operation of these subsystems. In addition, these systems produce log files to check instrument operation and register errors. Figure \ref{fig:sistema_controle_SPARC4} presents a diagram of the SPARC4 control system.

Before describing the software used in the SPARC4 operation, we introduce its two operational modes: photometric and polarimetric modes. In the photometric mode, no change in the instrumental configuration occurs along an acquisition series. In this mode, S4GUI sends a trigger command to each S4ACS instance to start the acquisition of an image sequence. The image acquisition in the polarimetric mode is more complex. Initially, S4GUI sends a request to S4ICS to put the waveplate on the first position of the sequence. Then S4GUI sends a request to the S4ACS instances to trigger the acquisition on the cameras. At the end of the exposures in all four channels, S4GUI sends a new request to S4ICS to move the waveplate to the next position. This process continues until the end of the polarimetric series. However, this process requires continuous communication among several components and, therefore, there might be some delays related to each one of these steps. These delays play a crucial role in the temporal resolution of the polarimetric mode of SPARC4. Table \ref{tab:operation_modes} presents a complete list of modes and sub-modes of SPARC4 operation.

\begin{table*}[tbh]
    \centering
    \caption{Operational modes of SPARC4.}
    \begin{tabular}{lccccc}
        \hline
        \hline
        Operational Mode & Analyzer & Half-Wave Plate & Quarter-Wave Plate & Polarizer & Depolarizer    \\ 
        \hline
        Photometry & $\dots$ & $\dots$ & $\dots$ & $\dots$ & $\dots$  \\
        Linear polarimetry & In the beam & In the beam & $\dots$ & $\dots$ & $\dots$ \\
        Linear polarimetry efficiency & In the beam & In the beam & $\dots$ & In the beam & $\dots$ \\
        Instrumental linear polarimetry & In the beam & In the beam & $\dots$ & $\dots$ & In the beam \\
        Circular polarimetry & In the beam & $\dots$ & In the beam & $\dots$ & $\dots$ \\
        Circular polarimetry efficiency & In the beam & $\dots$ & In the beam & In the beam & $\dots$ \\
        Instrumental circular polarimetry & In the beam & $\dots$ & In the beam & $\dots$ & In the beam \\
        \hline
    \end{tabular}
    \label{tab:operation_modes}
\end{table*}

SPARC4 has two modes for the synchronization of the scientific cameras acquisition: synchronous and asynchronous. In the asynchronous mode, the acquisition trigger is requested independently in each channel. The synchronous mode is the most used mode in scientific observations. In this mode, each new series of images from the four cameras is triggered simultaneously by S4GUI. Both of these synchronization modes can be used for photometric or polarimetric observations. However, polarimetric acquisitions using the asynchronous mode can be made for just one camera at a time.

\subsection{SPARC4 Acquisition Control System} \label{sec:ACS}

\begin{figure}[tbh]
    \centering
    \includegraphics[scale=0.45]{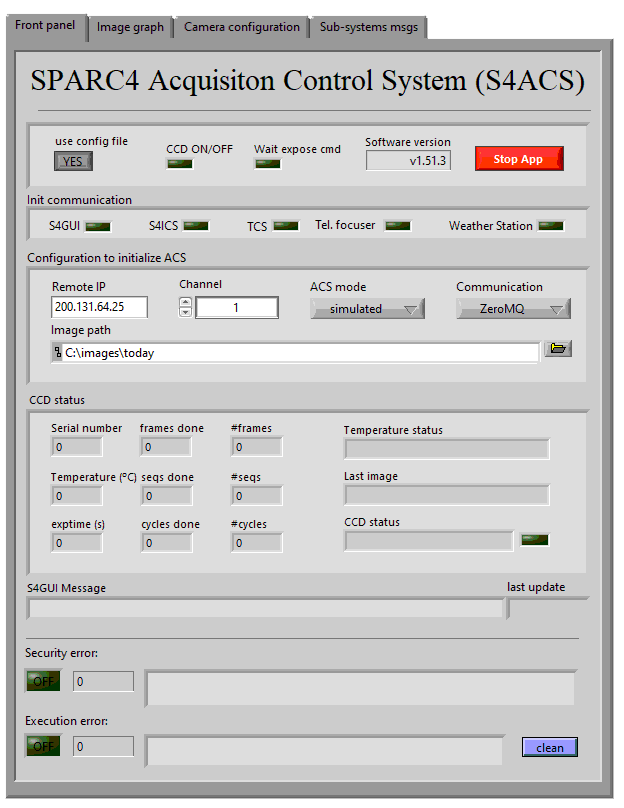}
    \caption{Front panel of S4ACS. This panel can be split into six sub-regions. From the top to bottom, the first one presents the controls of S4ACS as well as the status information related to the software execution, the second one presents the communication status between S4ACS and the other instrument subsystems and observatory facilities, the third one presents the configuration for initializing S4ACS, the fourth sub-region presents the current state of the camera, the fifth sub-region presents the content of the messages received from S4GUI, the sixth sub-region presents the error messages generated during the execution of S4ACS.}
    \label{fig:ACS}
\end{figure}

S4ACS \citep{Denis_SAB_2023,S4ACS_2024} is a software developed using the Laboratory Virtual Instrument Engineering Workbench (LabVIEW) 2018 programming language and the Software Development Kit (SDK) package, version 2.104.30000.0 \citep[02-24-2020, ][]{SDK}, made available by Oxford Instruments, to control the SPARC4 EMCCDs (see Figure \ref{fig:ACS}). The operating system used for the development of S4ACS was the Windows Server 2016 Standard, version 1607, and the methodology used for its development was based on \cite{Bernardes2019}. Using S4ACS, the parameters of the SPARC4 scientific detectors can be configured, and one or many series of images can be acquired. The acquired data are saved in a Flexible Image Transport System (FITS) file. All the information related to the acquisition, such as the instrument configuration and the telescope and weather information, are written into the FITS file header in parallel with the acquisition. Both the creation of the files and the edition of the headers are done using scripts developed in the Python language. These scripts are run using a Python interpreter version~3.6, running in the LabVIEW platform, using a native library to create Python sessions. With the current S4ACS version (v1.51), a series with 1400 images of 1024~$\times$~1024~pixels can be acquired, with an overhead of 4.5~ms between images. In addition, a sequence of series of 1400~images can be acquired, with an overhead of 91~ms between series, when using the asynchronous mode. However, the overhead in SPARC4 acquisition can be affected by other instrument subsystems. Section \ref{subsec:performance_control_system} shows the values for the different operational modes.

Each S4ACS instance communicates with SPARC4 and the observatory subsystems using the ZeroMQ\footnote{\url{https://labview-zmq.sourceforge.io/?home}} protocol. This is an open-source messaging library that can communicate with several different languages using Ethernet. S4ACS uses two communication patterns. In the first, it periodically publishes its status using the publish-subscribe pattern. S4GUI, in turn, uses these publications to access the status of the 4 channels at any time. Moreover, S4ACS uses this communication pattern to access the status of the instrument and the observatory subsystems, which are written into the image headers. The second pattern used by S4ACS is the request-reply. This pattern is exclusively used in receiving the requests sent by S4GUI, as a command to set the cameras configuration or even a trigger for starting the acquisition of an image series.

During the development of S4ACS, we verified that the SDK of scientific cameras is capable of communicating with only one camera at a time. For the control of several cameras, the current active device should be set. This procedure would reduce the performance of the instrument if more than one camera is controlled by the same computer. This occurs because the reading of the acquired data by two different cameras cannot be performed simultaneously. To avoid this problem, we decided to use one camera per computer and, for each computer, one instance of S4ACS. \cite{Piirola2021} found the same solution for the polarimeter Dipol-UF. 

S4ACS presents two operation modes: real and simulated. In real mode, S4ACS performs a real acquisition. Therefore, all the tasks performed by the software, as setting the cameras operation mode or reading the acquired image, occur using SDK to control the camera. In the simulated mode, on the other hand, S4ACS simulates all the functions that require the use of SDK. In simulated acquisitions, for example, S4ACS creates an image filled with zeros with the same size request by the user. This mode is useful when performing engineering tests on the system, avoiding the use of a real camera.

One of the acquisition parameters of the scientific Andor cameras is related to the number of exposures acquired for a single acquisition request. S4ACS limits this parameter to be chosen between single and kinetic. In single acquisition mode, one image is acquired for each acquisition request. In kinetic acquisition mode, many images can be acquired for one request. In the vendor Andor acquisition software, the kinetic mode produces an image cube: a single FITS file that has as many extensions as the number of exposures in the cube. S4ACS, on the other hand, saves each exposure in a single FITS file with a complete header. This processing is done during the acquisition of the kinetic series, so the FITS file corresponding to a given exposure in the series can be inspected about 148~ms after its acquisition and before the end of the kinetic series. The default mode for SPARC4 is kinetic, in which a series of up to 1400 images can be acquired.

The maximum number of exposures in an image cube in kinetic mode is limited by the internal buffer of the camera, which has a capacity of 3~GB, as determined using SDK. However, an issue has arisen when attempting to acquire multiple series of 1400~full-frame images using the camera fastest readout rate of 30~MHz. In this configuration, the acquisition rate exceeds the rate at which S4ACS can save images to the disk, causing a LabVIEW memory crash. Although a solution is still under development, the current performance of the instrument is sufficient to meet the majority of SPARC4 scientific requirements.

S4ACS provides a series of security procedures to prevent a misuse of the EMCCD cameras. One of them is related to the configuration of the camera parameters, which are described in \cite{Denis_2021}. The SPARC4 cameras present several parameters to control their operation. The proper choice of these parameters can be a little tricky without some technical knowledge. For this reason, S4ACS performs a sequence of verification steps over the parameter values provided by S4GUI, raising a warning in the case of any inconsistency. This feature is particularly important because S4ACS can be used by other softwares (nor only S4GUI), so S4ACS itself should guarantee a safe operation of the camera.

To help diagnose the behavior of S4ACS during its use, three files are created for each observation night. These files are used to log relevant information related to the operation of S4ACS. In the first file, unexpected errors that might occur while S4ACS is running are logged. The second file is used to log any inconsistency found between the information received from the observatory subsystems and the format used in SPARC4. In the last file, events considered important during the execution of S4ACS, such as requests received from S4GUI or incorrect incoming requests of camera configuration, are logged.

\subsection{Instrument Control System of SPARC4} \label{sec:S4ICS}

The SPARC4 Instrument Control System (S4ICS) is composed of S4ICSoft and the instrument mechanisms. The mechanisms correspond to the motors and electronic equipment used to accurately position the SPARC4 optical devices, according to the instrument configuration. S4ICSoft is the software responsible for controlling these mechanisms. In the following, each one of these components are described.

\subsubsection{Instrument mechanism components} \label{subsubsec:instrument_mech}

The mechanisms used to control the optical components of the instrument are composed of an Arcus DMX-ETH-17-2 NEMA 17 stepper motor, which provides a torque of 0.3~Nm \citep{Arcus}, coupled to precision worm wheel gearboxes, models PF20-10NM, P20-60A and PF20-60ANM, manufactured by Ondrives.US. These motors are equipped with an integrated encoder of 1000 pulses per revolution (ppr), a hardware driver, I/O lines for triggering external events, and a communication interface based on the Transmission Control Protocol/Internet Protocol (TCP/IP). Up to two standalone custom programs can be uploaded and executed into the memory of these devices, allowing their control via commands sent via the TCP/IP interface. The stepper motor used in the autoguider focuser mechanism is a NPM PF35T-48C4G model, of 35~mm in diameter, with a torque of 24~Nm, and 48~ppr and is the only one that differs from the others. An electronic circuit was built to control this motor using the I/O signals of the patrol mirror stepper motor. With this solution, the focuser can be controlled in the same way as used for the other motors, standardizing the way in which these mechanisms are managed. 

\subsubsection{S4ICSoft} \label{subsubsec:S4ICSoft}

S4ICSoft, currently in version 20240924, is a software developed in LabVIEW 2018, designed to control all the moving elements of SPARC4. This control is done using three routines saved inside the stepper motor memories: BOOT, INIT, and GOTO. The BOOT program identifies which optical component is attached to the stepper motor and runs automatically when the motor is energized. 
The INIT and GOTO routines are only executed using commands sent via the TCP/IP interface. The INIT routine loads several parameters related to the mechanism (as its offset, transmission ratio, and speed) and performs a calibration of the motor enconders. This procedure is important due to small errors related to the positioning of these motors during their operation.
The GOTO routine manages the direction of rotation, speed, and sensor readings to move the optical component to the desired position. The BOOT routine is common to all the mechanisms, whilst the INIT and GOTO routines were specifically developed for each one of them. The set of all these routines makes up the called S4DMX package, which works like a firmware for the motors. The source code for both S4DMX\footnote{\url{https://github.com/fco-R/S4DMX}} and S4ICS\footnote{\url{https://github.com/fco-R/S4ICS}} can be found on the GitHub platform. Similarly to S4ACS, S4ICSoft provides a simulated mode for these mechanisms. In this mode, all tasks related to the operation of these stepper motors are simulated programmatically. This option was developed for engineering purposes and allows to test S4ICS without the real devices.

\begin{figure*}[tbh]
    \centering
    \includegraphics[width=\textwidth]{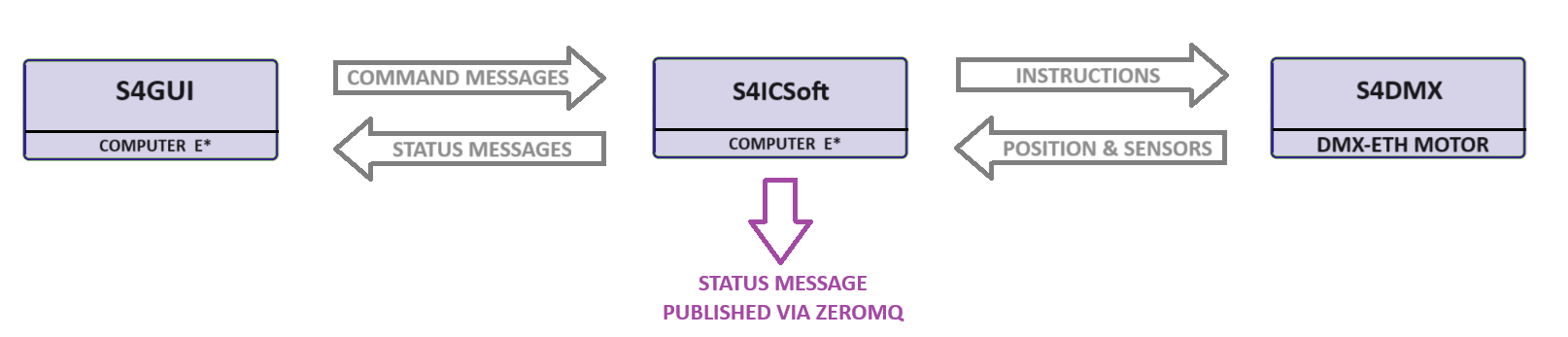}
    \caption{Flux of information in a communication between S4GUI, S4ICSoft, and an instrument mechanism. In the figure, the status message published by S4ICSoft via the ZeroMQ package is also presented.}
    \label{fig:S4ICS_Communication}
\end{figure*}

To control these mechanisms, S4GUI communicates with S4ICSoft using the ethernet network. For this communication, S4ICS provides three different methods: TCP/IP, ZeroMQ, or Message Queuing Telemetry Transport (MQTT). Upon receiving a request from S4GUI, S4ICSoft verifies its validity and, if valid, sends the instruction to the motor. During the same time, S4ICSoft continuously monitors the status of each individual mechanism and relays this information to S4GUI to ensure proper instrument control. Figure \ref{fig:S4ICS_Communication} presents a diagram of the information flux between these components. Before establishing communication, the internal S4ICSoft TCP/IP server must be started. This can be done automatically upon launching the software or manually by the user, which requires a login. Once the server is running, S4ICSoft loads its last configuration, and the control panel displays the current position and status of the mechanisms.

The SPARC4 Communication Library (S4CL) is a custom language specifically developed for the communication between S4GUI and S4ICS, which supports two types of messages: command messages and status messages. A command message begins with the prefix “S4ICS”, followed by the name of the mechanism and the requested action, which can be one of the following: INIT, GOTO, or STOP. The GOTO action requires an additional value specifying the target position of the mechanism, which can be a numerical value (converted into motor encoder units by S4ICSoft) or a predefined position name. For example, the values of ``L/2", ``L/4", or ``OFF" can be used for the waveplate selector. In contrast, the INIT and STOP commands are followed by the “NONE” statement. S4ICSoft also allows to concatenate multiple commands for different mechanisms within the same request.

The status messages are composed of concatenated statuses of individual mechanisms. Each mechanism status includes the mechanism name, mode, condition, position value, position name, and position ID. The mode can have one of three values: ACTIVE, SIMULATED, or NONE. ACTIVE and SIMULATED values indicate if S4ICSoft is controlling a real or a simulated mechanism. The NONE value is used when the motor is disabled or has no communication. The condition field represents the status of the mechanism and can be set to READY, BUSY, TIMEOUT, or NONE. TIMEOUT is used when a previous command could not be completed. The position value shows the last position of the mechanism within its operational range. When a predefined position is used, the position name and ID fields provide the corresponding name and index. Otherwise, these fields are set to NONE and -1. S4ICS continuously publishes status messages of the mechanisms via ZeroMQ. These statuses are updated in real-time when a mechanism is executing a task, and every three seconds when the mechanisms are idle.

\begin{figure*}[tbh]
    \centering
    \includegraphics[width=\textwidth]{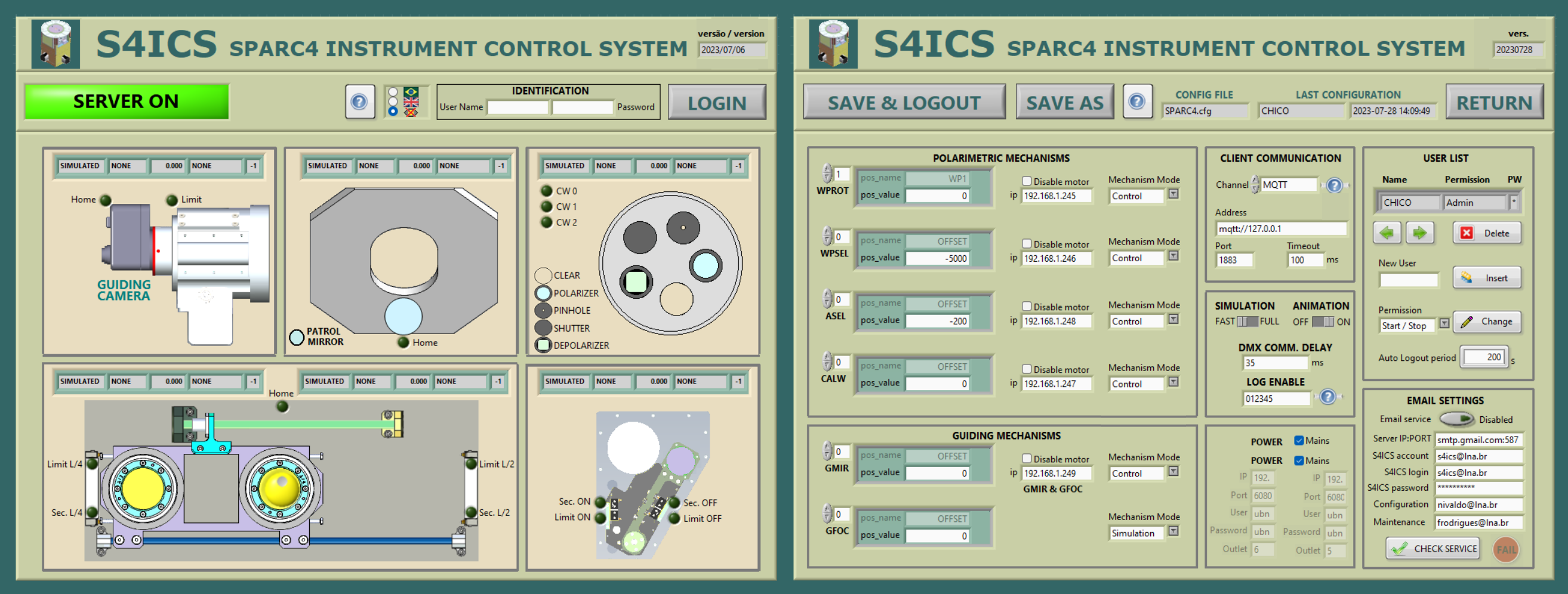}
    \caption{S4ICSoft: control panel (left side) and configuration panel (right side).}
    \label{fig:S4ICS_Panels}
\end{figure*}

Figure \ref{fig:S4ICS_Panels} presents the control panel (left side) and the configuration panel (right side) of S4ICSoft. The control panel provides real-time information about each mechanism status and position, while the configuration panel is reserved for engineering purposes and can only be accessed by users with special permissions. The configuration panel is divided into several sub-panels. The panels of the Polarimetric Mechanisms and the Guiding Mechanisms can be used to control the mode (simulated or real), set the IP addresses, and enable or disable each mechanism. This panel also allows for manual control of their position. The Client Communication panel is used to choose between the communication protocols provided by S4ICSoft and to set the IP address in which the status messages are published. The panel of the User List is used to manage the authorized users login accounts and permissions, in need of any manual intervention. The Email Settings panel allows users to configure email notifications that alert the observatory when an error is detected in the software operation. The remaining panels are the General Utilities and Power Management. General Utilities is used for configuring general options related to the S4ICSoft operation, like starting the remote server automatically, or selecting which level of events generated during the software execution are logged. The Power Management frame is not implemented yet, but will be used to turn on and off the power supplies of equipment and motors.

\subsubsection{Mechanism performance} \label{mechamism performance}

In this section, the main parameters of the S4ICS mechanisms considered relevant for the SPARC4 control system are presented. These parameters were measured in laboratory tests, in which the stepper motors were exposed to temperatures ranging from -5~$^{\circ}$C to 40~$^{\circ}$C in a thermal chamber. In addition, these components were tested using the standalone mode. With this configuration, the timings related to the movement of these mechanisms were obtained, avoiding contributions of the communication between S4ICS and S4GUI. The results of these tests are presented in Table~\ref{tab:mechanisms_time_charact}. Following the order of the columns presented, the operation range is the first parameter and represents the mechanical operation range in mm, for linear mechanisms, and in degrees, for rotatory mechanisms. The step resolution is the correspondence between the encoder resolution (1000 ppr) and the minimal mechanical displacement of each mechanism. The initialization time is the time needed for each mechanism to move between its home position and a predefined reference position. This parameter refers to the time required by the control system to switch between photometric and polarimetric modes. During this transition, each optical component is placed in position simultaneously. As a result, the total time needed for the switch is determined by the waveplate, which is the slowest optical component. The procedure of placing or removing this component from the light path consists of a combination of the waveplate selector, followed by the waveplate rotator mechanism movements. The positioning time represents the time it takes for each mechanism to move to the next position. The waveplate is rotated between images in a polarimetric sequence. Therefore, its positioning time is important to determine the polarimetric acquisition rate provided by the instrument, as described in Section \ref{subsec:performance_control_system}. Positioning precision is the maximum error provided by the motor when placing the optical component in the light path.

\begin{table*}[tbh]
\caption{Main parameters of the S4ICS mechanisms, measured in laboratory tests. The columns show the name of the mechanism, its mechanical operation range, positioning resolution, initialization and positioning times, and precision.}
\begin{tabular}{cccccc}
\hline
\hline
Mechanism     & Operation    & Step       & Init. Time & Pos. Time & Positioning  \\ 
              & Range        & Resolution & (s)        & (s)       & Precision \\
\hline
Waveplate rotator    & 360$^{\circ}$ & 43.2 arcsec & 6           & 0.3     & 43.2 arcsec \\
Waveplate selector   & 160 mm        & 0.2 $\mu$m  & 100         & 45      &  0.5 $\mu$m      \\
Calibration wheel    & 360$^{\circ}$ & 21.6 arcsec & 25          & 1.5     &  10  $\mu$m      \\
Analyzer      & 30$^{\circ}$  & 21.6 arcsec & 30          & 5       &  20  $\mu$m      \\
Patrol mirror & 360$^{\circ}$ & 21.6 arcsec & 17          & 2       &  21.6 arcsec \\
Focuser       & 19 mm         & 12.2 $\mu$m & 65          & 33      &  18 $\mu$m    \\ 
\hline   
\end{tabular}
\label{tab:mechanisms_time_charact}
\end{table*}

\subsection{Graphical User Interface of SPARC4} \label{sec:S4GUI}

S4GUI is a graphical user interface developed using the LabVIEW 2018 programming language and hosted on the GitHub platform\footnote{\url{https://github.com/overducci/SPARC4_GUI}}. It communicates with S4ICS and S4ACS, allowing for the operation of the instrument and the acquisition of images with a high level of automation. Besides, S4GUI also communicates with observatory facilities, as the Telescope Control System (TCSPD) and the Weather Station, to get information about the observation conditions. A software description on the architecture and the execution flow for S4GUI is presented in Appendix \ref{sec:appendixA}.

Figure \ref{fig:gui} presents the main front panel of S4GUI version 20241016, which allows the user to control the instrument functions and monitor the observation conditions. Following the regions highlighted in the figure, region A presents the communication status between S4GUI and the other instrument subsystems as well as the observatory facilities. In addition, this region also presents the current telemetry information. Region B presents the trigger mode and the current state of the acquisition. Region C presents the tabs for the predefined observational setups. Region D presents the instrument mode and the polarimetry setup. Region E presents the controls of the instrument channels. Region F presents the acquisition controls. Region G presents the current status of the cameras. In the next sections, the main aspects related to the S4GUI operation are described.

\begin{figure*}[tbh]
    \centering
    \includegraphics[width=\textwidth]{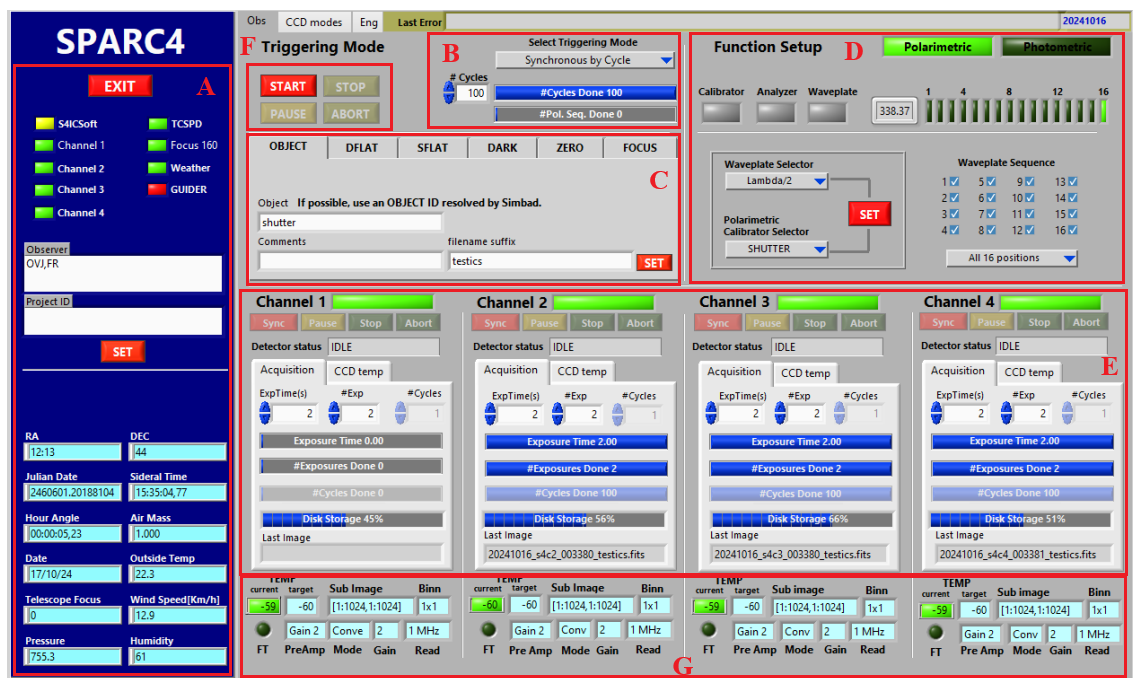}
    \caption{Main front panel of S4GUI. In the figure, the main groups of tasks performed by this software are highlighted in specific regions as follows. A: observatory telemetry and connection status; B: region for selecting the triggering mode; C: predefined CCD and image settings; D: instrument mode setup and status; E: controls of the instrument channels; F: acquisition controls; G: indicator of the camera current statuses.}   \label{fig:gui}
\end{figure*}

\subsubsection{SPARC4 operation modes} \label{operation modes}

SPARC4 provides the polarimetric and photometric operation modes (Table~\ref{tab:operation_modes}). For the polarimetric mode, the waveplate (half-wave or quarter-wave) and the sequence of waveplate positions should be set. For waveplate positions, up to sixteen values varying from 0\degr\ to 337.5\degr\ equally spaced by 22.5\degr, can be freely selected. The selection of these positions can be done by clicking the check boxes or selecting one of the preconfigured options (all 16; first 8; first 4; 1 to 4 plus 9-12) provided by a drop-down menu. For instrument characterization purposes, S4GUI also provides the option of choosing the position of the calibration wheel.

\subsubsection{Acquisition configuration}

\begin{figure}[tbh]
    \centering
    \includegraphics[width=\columnwidth]{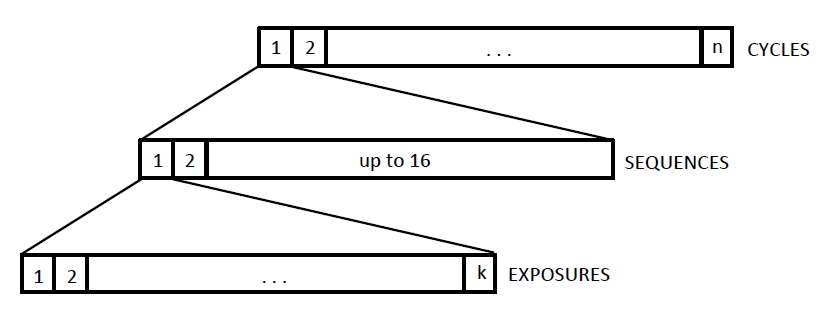}
    \caption{Relationship of cycles, sequences, and exposures for an acquisition of an image series. In photometric mode, the number of sequences is always one.}
    \label{fig:cycles}
\end{figure}

The SPARC4 acquisition configuration is based on the concepts of sequences and cycles, which are best understood in polarimetric mode. An acquisition requires the choice of the exposure time and the number of exposures for each waveplate position. These values should be set for each channel and are not necessarily the same. This set of exposures for one waveplate position is named a sequence. A cycle consists of one set of sequences that includes all selected waveplate positions. An acquisition can be made up of any number of cycles, and its generic structure is represented in Figure \ref{fig:cycles}. Therefore, for a given channel, a choice of 3 cycles of 16 waveplate positions (sequences) of 10 exposures corresponds to the acquisition of 3~$\times$~16~$\times$~10~=~480 images. In the synchronous polarimetric mode, the number of cycles and the number of sequences (waveplate positions) in a cycle should be the same for all channels because the exposures in a given sequence should be acquired with a specific retarder waveplate position, and the beam passes through the retarder before the split in channels. The START button triggers the acquisition on all channels. Asynchronous polarimetric acquisitions can be performed for a single channel at a time.

In the photometric mode, the number of sequences is always one, by definition. The synchronous mode adopts the same number of cycles for all channels, and the START button triggers the acquisition in all channels, as in the polarimetric mode. However, contrary to the polarimetric mode, the photometric mode can operate asynchronously or synchronously for any number of channels because the images in each channel can be acquired independently. Therefore, in asynchronous photometric mode, each channel must be set to its own number of cycles, and its acquisition can be individually controlled using the START, PAUSE, STOP and ABORT buttons, present in their specific tabs (see region E of Figure \ref{fig:gui}). 

\subsubsection{Predefined observational setups}

\begin{figure}[tbp]
    \centering
    \includegraphics[width=\columnwidth]{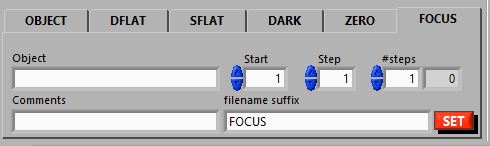}
    \caption{Tab for the predefined settings for the automatic focus adjustment. For that, the user should provide the start value, the step size for the next position, and the number of steps to be done for moving the telescope focus during the image acquisition.}
    \label{fig:focustab}
\end{figure}

S4GUI provides a set of predefined observational setups (region C of Figure \ref{fig:gui}) to assist the user with commonly used observational procedures. The choice of the OBJECT tab automatically sets the OBSTYPE header keyword to OBJECT. The dome flat (DFLAT) and sky flat (SFLAT) tabs set the filename suffix to “dflat” and ``sflat", respectively, and the OBSTYPE keyword to FLAT. 
The DARK tab sets the camera shutter to permanently closed, the OBSTYPE keyword to DARK, and the filename suffix to “dark”. The ZERO tab permanently closes the camera shutter, switches the FT to off, and fixes the exposure time in $1 \times 10^{-5}$~s. For this tab, the OBSTYPE keyword is ZERO. Leaving the ZERO tab, the FT and the exposure time return to the previous used settings. Finally, the FOCUS tab assists the user in obtaining the telescope focus for the used instrument configuration and observation conditions. This is done by automatically acquiring a series of images for different positions of the secondary mirror of the telescope. For that, the user should provide the values for the start, step, and number of steps for the telescope focus (see Figure \ref{fig:focustab}). When pressing the START button, S4GUI starts the acquisition of a series of images based on the values provided. The obtained series can then be used by an external application to calculate the best focus value. The OBSTYPE keyword and suffix are set to FOCUS and “focus”, respectively. Table \ref{tab:predefined_CCD_setup} summarizes these predefined configurations.

\begin{table*}[tbh]
    \centering
    \caption{Predefined observational setups. The columns present the tab name, values for the OBSTYPE keyword, object name, filename suffix, shutter, frame transfer, and exposure time. In the last two columns, PS stands for the previous setting used in an acquisition, before choosing the respective tab.}
    \begin{tabular}{ccccccc}
        \hline
        \hline
        Tab & OBSTYPE & Object & Filename & Shutter & Frame  & Exposure Time \\
        & Keyword & Name & Suffix & & Transfer & (s)\\
        \hline
        OBJECT & OBJECT  & $\dots$ & $\dots$ & Open & PS & PS\\
        DFLAT  & FLAT    & DOMEFLAT     & dflat & Open & PS & PS\\
        SFLAT  & FLAT    & SKYFLAT     & sflat & Open & PS & PS\\
        DARK   & DARK    & $\dots$ & dark & Closed & PS & PS\\
        ZERO   & ZERO    & ZERO         & zero & Closed & Off & 1E-5\\
        FOCUS  & FOCUS   & $\dots$ & focus & Open & PS & PS\\
        \hline
    \end{tabular}
    \label{tab:predefined_CCD_setup}
\end{table*}

\newpage
\subsubsection{Configuring the CCD operation modes}

\begin{figure*}[tbh]
    \centering
    \includegraphics[scale=0.5]{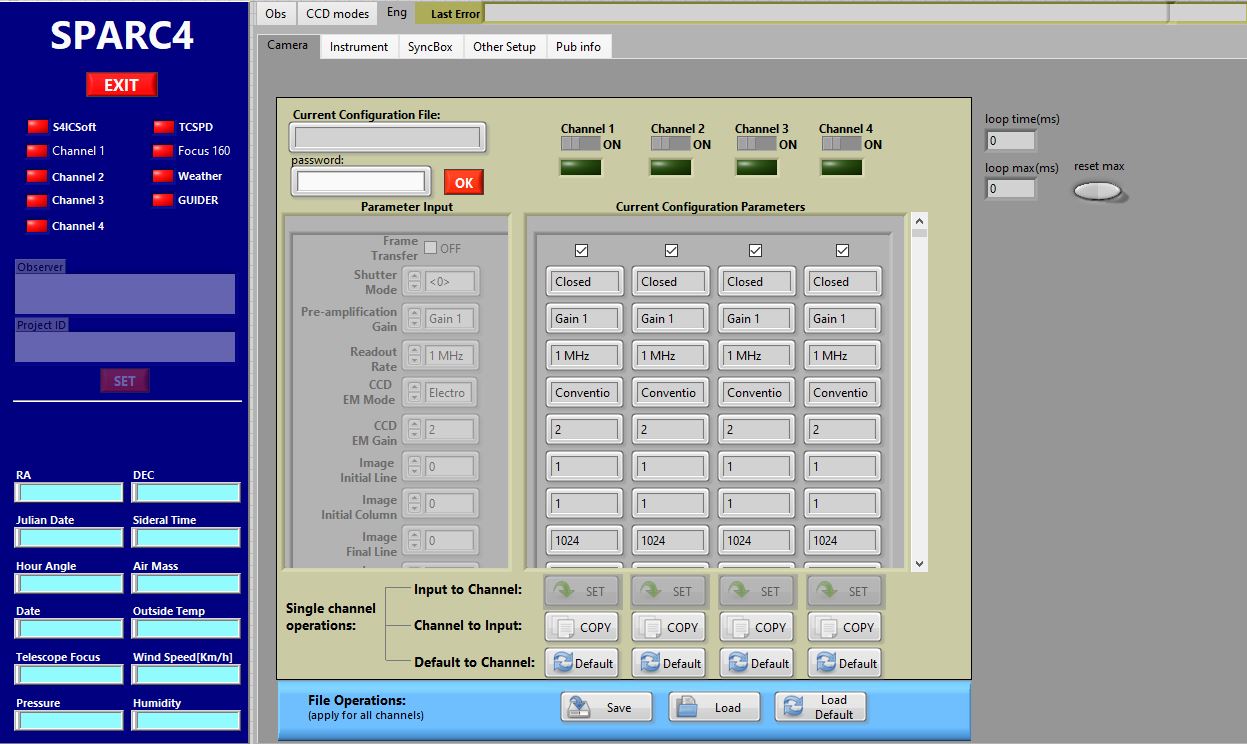}
    \caption{Camera panel for configuring the cameras operation mode. This panel can be accessed in the Engineering tab, presented on the top of the figure.}
    \label{fig:guiccd}
\end{figure*}

The Camera panel can be accessed in the Eng tab, presented at the top of Figure \ref{fig:guiccd}. This panel is used to configure the operation mode of the SPARC4 cameras as well as to enable or disable the devices that should be used during the observation. In setting an operation mode, the values for the parameters presented in the ``Parameter input" column should be configured. After configuring the operation mode, the SET button should be pressed for the corresponding channel. However, both the Eng tab and the camera parameters are password-protected and can be accessed only by authorized personnel. For configuring the operation mode of the cameras, the CCD modes tab presented in Section~\ref{subsec:ccd_modes} should be used.

\subsubsection{CCD Modes} \label{subsec:ccd_modes}

\begin{figure*}[tbh]
    \centering
    \includegraphics[scale=0.5]{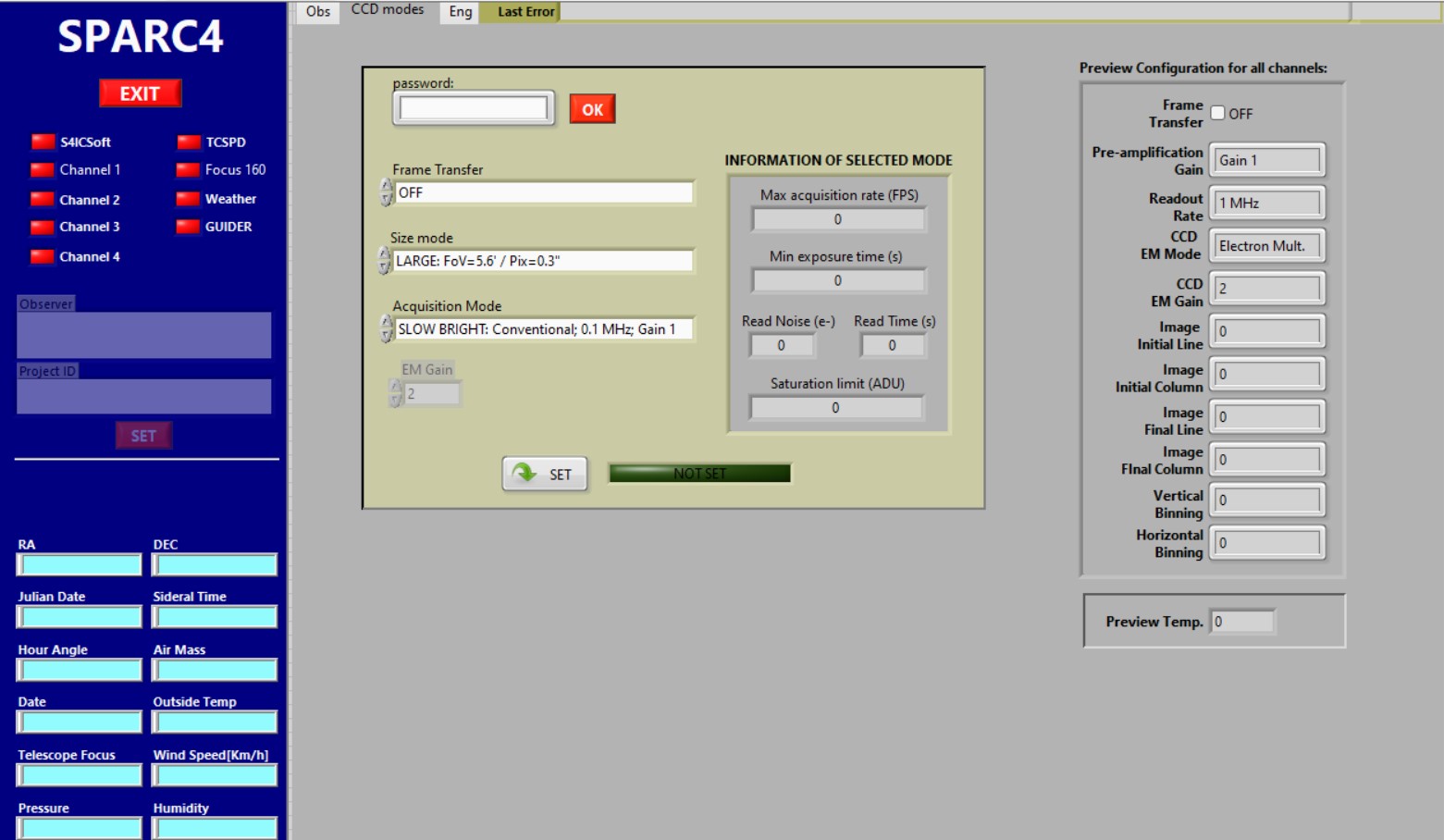}
    \caption{CCD modes panel for choosing the detectors operation mode.}
    \label{fig:ccdmodes}
\end{figure*}

The CCD mode tab shown in Figure \ref{fig:ccdmodes} provides a series of options to assist the user in choosing a configuration for the SPARC4 detectors. There are three parameters for this: the FT mode, the size mode, and the acquisition mode. The FT mode allows one to enable or disable the FT mode. The acquisition mode comprises the output amplifier, readout rate, and preamplification gain to adjust the sensibility and readout rate. The size mode allows one to configure the FoV and pixel binning. A summary of the options provided by the acquisition mode and the size mode are presented in Table \ref{tab:acq_modes}. When these parameters are configured, the maximum acquisition rate, minimum exposure time, read noise, readout time, and saturation limit related to the selected mode are presented. For the provided modes, the image readout time are in 5.9~ms to 1.24~s range. By pressing the SET button, this mode is transferred to the cameras, and an LED informs the user if the new camera setup was successfully configured.

\begin{table*}[tbh]
    \centering
    \caption{The standard acquisition and size modes of SPARC4. For the acquisition modes, the columns show the corresponding values for the output amplifier, readout rate, and preamplification gain. The acronym ``EM" denotes the Electron Multiplying option of the output amplifier. Regarding the size mode, the columns provide information on the Field of View (FoV), centered on the camera chip, and pixel binning, which refers to the number of pixels processed by the camera electronics as a single pixel. For both modes, the maximum acquisition rate attainable by the respective configuration is also presented.}
    \begin{tabular}{ccccc}
        \hline
        \hline
         Acquisition Mode & Maximum & Output & Readout Rate & Preamp Gain \\
          & Acq. Rate (fps) & Amplifier & (MHz) & \\
        \hline
        SLOW and BRIGHT & 0.8 & Conventional & 0.1 & 1 \\
        SLOW and FAINT  & 0.8 & Conventional & 0.1 & 2 \\
        NORMAL and BRIGHT & 7.14 & Conventional & 1 & 1 \\
        NORMAL and FAINT  & 7.14 & Conventional & 1 & 2 \\
        FAST & 66.67 & EM & 10 & 1\\
        FAST+ & 120.48 & EM & 20 & 1\\
        SUPERFAST & 169.49 & EM & 30 & 1 \\
        \hline
        Size Mode & Maximum & FoV & Binning\\
         & Acq. Rate (fps) & (arcmin) & (pixels) \\
        \hline
        LARGE  & 25.64 & 5.6 $\times$ 5.6 & 1 $\times$ 1 \\
        LARGE  & 50.00 & 5.6 $\times$ 5.6 & 2 $\times$ 2 \\
        MEDIUM & 50.00 & 2.8 $\times$ 2.8 & 1 $\times$ 1 \\
        MEDIUM & 90.91 & 2.8 $\times$ 2.8 & 2 $\times$ 2 \\
        SMALL  & 90.91 & 1.4 $\times$ 1.4 & 1 $\times$ 1 \\
        SMALL  & 169.49 & 1.4 $\times$ 1.4 & 2 $\times$ 2 \\
        \hline
    \end{tabular}
    \label{tab:acq_modes}
\end{table*}

\subsubsection{Guider Setup}

The auto-guiding during SPARC4 acquisitions is based on two subsystems. The autoguider hardware is in the SPARC4 structure and its moving parts are controlled by SPARC4, specifically by S4ICS, as described in Section~\ref{sec:S4ICS}. The analysis of the autoguider images and the commands to the telescope pointing corrections are performed by the observatory software AGS. In this section, we describe the panel that allows the user to control the auto-guiding hardware. 

\begin{figure}[tbh]
    \centering
    \includegraphics[scale=0.5]{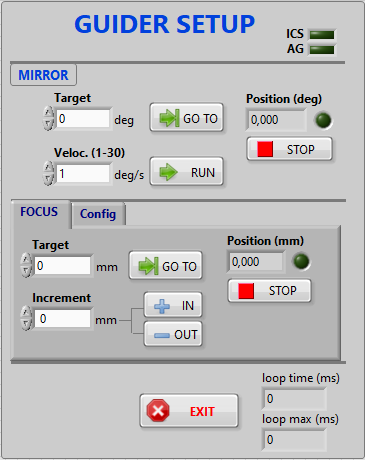}
    \caption{Front panel of the Guider Setup, used to control the patrol mirror and the telescope focus.}
    \label{fig:guidersetup}
\end{figure}

The Guider Setup (see Figure \ref{fig:guidersetup}) is a module within S4GUI designed to control the positions of both the autoguider focus and the patrol mirror. Patrol mirror movement can be performed by choosing a target position or velocity and then pressing the GO TO or RUN buttons, respectively. The focus position can be set to a defined target value using the GO TO button or by positive or negative steps with the IN and OUT buttons. Any movement can be stopped with the respective STOP buttons. There are two light indicators that turn on when the S4ICSoft and AGS connections are active.

The position of the patrol mirror defines the location in the sky of a guiding star. AGS requests, via TCP/IP, the position of the SPARC4 patrol mirror to determine the sky orientation. So, the Guider Setup module is responsible for replying the mirror status to the Autoguider whenever it is requested. This module also automatically sends focus movement commands to S4ICSoft according to an editable focus table that contains the best focus values for all instrument functions: photometry, polarimetry, polarimetry efficiency, and instrumental polarization. When a manual focus adjustment is made with the GO TO, IN, or OUT buttons, the Guide Setup will ignore the focus table until any change in the instrument function is detected.

\begin{figure}[tbh]
    \centering
    \includegraphics[scale=0.35]{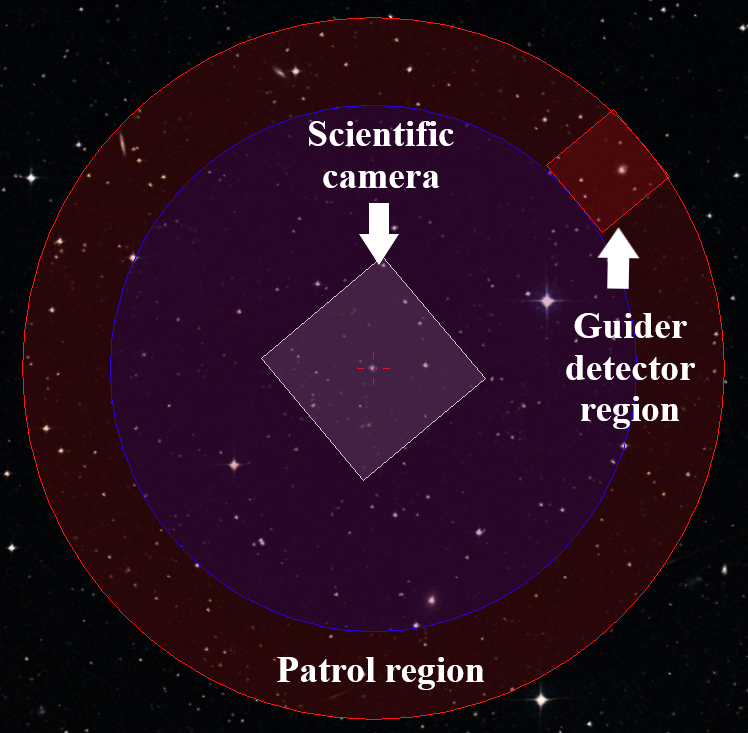}
    \caption{Field of the star GD 108 as viewed in Aladin Desktop \citep{aladin_desktop}. Over this field, the FoV obtained for the scientific camera, patrol region and guider detector region are presented. For this configuration, the chosen angle for the patrol mirror was 130$^{\circ}$.}
    \label{fig:aladin_sparc4_fov}
\end{figure}

To assist in the process of finding a guiding star, the Aladin Desktop software \citep{aladin_desktop} can be used. Aladin is an interactive sky atlas that integrates data from multiple astronomical catalogs. Using Aladin alongside the dimensions, in arcseconds, of the SPARC4 scientific camera, patrol region, and guider detector region, the FoV of these components in the sky can be determined.  Figure~\ref{fig:aladin_sparc4_fov} provides an example of this scenario for the spectrophotometric standard star GD~108 \citep{Oke_1990}. With this configuration, the patrol mirror position angle that should be used for the guiding star can be found by adjusting the position of this component in the Aladin field viewer.

\newpage

\subsection{Acquisition overheads} \label{subsec:performance_control_system}

An important factor in astronomical time-series observations is the dead time between image acquisitions. This dead time is the time interval in which the cameras are not exposing and it dictates the temporal resolution of the instrument. In this section, we present the overheads related to the control system in the photometric and polarimetric modes of SPARC4. These overheads are the time required by the control system to perform a determined task during an acquisition. These overheads, together with the image readout time, compound the instrument dead time. To determine these overheads, several acquisitions were performed using full-frame images, a readout rate of 1~MHz, and the kinetic and FT modes of EMCCDs. According to the manufacturer, the readout time for this acquisition mode is 1.11~s. For simplification, the results presented in this section are related to the channel g of SPARC4. However, the experiments showed that these overhead values are the same for all channels within the errors.

\subsubsection{Photometric mode} \label{subsec:overhead_phot_mode}

In photometric mode, the overhead manifests in two forms: overhead between exposures and overhead between cycles. The overhead between exposures was evaluated by acquiring two series of 100 exposures in a single sequence and a single cycle, one with an exposure time shorter than the readout time ($t_{exp} < t_r$), and another with an exposure time longer than the readout time ($t_{exp} > t_r$). The overhead was calculated by the difference between each two subsequent exposures in the series. For measurements where $t_{exp} < t_r$, the used exposure time was 1~$\times$~10$^{-5}$~s and the average value obtained for the overhead was 1.1099078~$\pm$~5~$\times$~10$^{-7}$~s. This value is the limit of the acquisition frequency for this acquisition setup \citep{SDK} and is equal to the readout time presented by \cite{Denis_2021}, considering the margin of error. Since this time is determined by the camera electronics, an increase in its value was unexpected. For the measurement where $t_{exp} > t_r$, an exposure time of 2~s was used and the overhead was 4.4958~$\pm$~4~$\times$~10$^{-4}$~ms. In this case, $t_{exp}$ is the limit of the acquisition frequency and the value found for the overhead corresponds to the theoretical value of 4.43~ms, suggested by the camera hardware manual. This is the time to transfer the charges from the imaging region to the storage region of the CCD, in FT mode \citep{ixon_hardware_guide}. 

To measure the overhead between cycles, 21 cycles of 1~exposure were acquired using an exposure time of 2~s, for the synchronous and asynchronous modes of the instrument. The overhead was calculated by the difference between the times registered in the header for each two subsequent images. This result was subtracted from the exposure time and the readout time of 1.11~s. The overhead obtained for these experiments were 91 $\pm$ 2~ms and 453 $\pm$ 54~ms for the asynchronous and synchronous modes, respectively. In asynchronous mode, the start of the next cycle is done directly by S4ACS and about 18.6~ms of this time corresponds to the time required by SDK to start a new acquisition by the cameras. In the synchronous mode, in turn, S4GUI needs to identify the end of the current cycle in all channels before sending an expose command to the S4ACS instances start the next cycle. Therefore, the difference in the overhead between these cases is given by the time that S4GUI needs to communicate with all S4ACS instances.

\subsubsection{Polarimetric mode}

In the polarimetric mode, the overhead can occur between exposures, sequences, and/or cycles. The overhead between exposures is exactly the same as that presented in the photometric mode, given that it is managed by the hardware of the cameras. Therefore, we discuss here only the overheads between sequences and cycles.

To evaluate the overheads, 11 cycles composed of 16 sequences, one for each position of the waveplate, were acquired. Each sequence, in turn, was composed of one exposure of 2~s of exposure time. The overhead between sequences is given by the difference between the time recorded in the image header of two subsequent exposures. The overhead between cycles was calculated by the difference between the time of the image headers for the last image of one cycle and the first image of the next cycle. Both of these results were subtracted from the exposure time and the readout time of 1.11~s. The overhead for sequences and cycles were 1.41 $\pm$ 0.11~s and 1.40 $\pm$ 0.12~s, respectively. These values represent the time needed for S4GUI controls S4ICS to move the waveplate to the next position, and send an expose command to the S4ACS instances. Given the configuration of the experiment, there is no difference in the time it takes for S4ICS to change the position of the waveplate between subsequent sequences and subsequent cycles. Consequently, the overhead values for sequences and cycles being similar within the margin of error aligns with this expectation.

The results obtained with these experiments are summarized in Table \ref{tab:overhead_values}. Based on these results and given the maximum number of images per sequence allowed by SPARC4, one can acquire a photometric time series of 1400~images with a minimum exposure time of 1.2~s, with an overhead of 4.5~ms between exposures, for the used CCD mode. 
For shorter exposure times, this same result can be achieved using faster readout rates. Also, SPARC4 allows to concatenate two cycles of 1400~images with an overhead of 91~ms between them, for the asynchronous acquisition mode. For the polarimetric mode, one cycle of 16 sequences can be acquired, with an overhead of 1.41~s between sequences. Moreover, polarimetric cycles can be concatenated with an overhead of 1.4~s between cycles. Therefore, SPARC4 needs a minimum time interval of about 41.8~s for the measurement of all the 16 positions of the waveplate, for the CCD used mode. For the cases of polarimetric measurements using 8 or 4 positions of the waveplate, this time interval is 20.9~s and 10.45~s, respectively. Similarly to the photometric case, using faster readout rates reduces the time required by SPARC4 to complete a polarimetric measurement. The best time resolution of a polarimetric measurement is defined by the sum of the exposure time, readout time, and the overhead to position the waveplate (1.41~s). Adopting specific polarimetric reduction procedures, this time resolution can be achieved in polarimetric time series.

\begin{table*}[tbh]
    \centering
    \caption{Overhead times from the control system obtained using different SPARC4 modes.}
    \begin{tabular}{cccc}
         \hline
         \hline
         Instrument & Synchronization & Measurement & Overhead\\
         Mode & Mode & Type & (s)\\
         \hline
         Photometric/polarimetric & $\dots$ & Images & 0.0044958 $\pm$ 4e-7\\
         Photometric& Asynchronous & Cycles & 0.091 $\pm$ 0.002 \\
         & Synchronous & Cycles & 0.45 $\pm$ 0.05 \\
         Polarimetric & Synchronous & Sequences & 1.41 $\pm$ 0.11\\
         & Synchronous & Cycles & 1.40 $\pm$ 0.12 \\
         \hline
    \end{tabular}
    \label{tab:overhead_values}
\end{table*}


\section{Conclusions} \label{sec:conclusion}

SPARC4 is a new instrument in use at the 1.6-m telescope of the Brazilian Pico dos Dias Observatory. It can operate as an imager or as a polarimeter and in both modes it acquires images in the SDSS g, r, i, and z bands simultaneously. In this work, the hardware and software of the SPARC4 control system is described. The hardware is composed of scientific cameras and devices related to the movements and positioning of the autoguider system and the polarimetric components. The three main softwares used to operate SPARC4 are S4ACS, S4ICS, and S4GUI. S4ACS is responsible for communicating and controlling the four scientific cameras. S4ICS controls the moving parts of SPARC4. Finally, S4GUI is the main user interface that orchestrates the SPARC4 subsystems.

For operations in the photometric mode, SPARC4 allows the acquisition of a cycle composed of up to 1400 images, with an deadtime of about 4.5~ms between images. Besides, several cycles of 1400 images can be concatenated with an overhead of 450~ms plus the readout time of the last image. For polarimetric acquisitions, a cycle of 16 sequences can be acquired with a deadtime of 1.41~s plus the image readout time between sequences. For both photometric and polarimetric modes, the readout time of the main acquisition modes is in the 5.9~ms~-~1.24~s range. This time, together with the overheads presented for each instrument mode, defines the temporal performance of SPARC4. In addition, the time to change from photometric to polarimetric mode (and vice versa) is a few minutes.

SPARC4 was developed with the aim of benefit a broad range of scientific cases. It is an ease to use instrument and its operational and acquisition modes can be changed quickly. It was highly demanded in the last OPD time-request call and we anticipate an important contribution to the observatory science production.


\section*{Acknowledgments}

SPARC4 was funded by Financiadora de Estudos e Projetos - Finep (Proc: 0/1/16/0076/00), Agência Espacial Brasileira - AEB (PO 20VB.0009), Fundação de Amparo à Pesquisa do Estado de São Paulo - FAPESP (Grant 2010/01584-8), Fundação de Amparo à Pesquisa do Estado de Minas Gerais - FAPEMIG (APQ-00193-15 \& APQ-02423-21), Conselho Nacional de Desenvolvimento Científico e Tecnológico - CNPq (Grant \#420812/2018-0) and INCT-Astrofísica. CVR also thanks CNPq (Proc: 310930/2021-9). DVB thanks to Coordenação de Aperfeiçoamento de Pessoal de Nível Superior - CAPES (Proc: 88887.513623/2020-00) for the scholarship funding. LF acknowledges the research productivity grant (PQ) number 316373/2023-0 awarded by CNPq. ACM thanks to CNPq (Proc: 150737/2024-6) for the fellowship funding. E.M. acknowledges funding from FAPEMIG under project number APQ-02493-22 and a research productivity grant number 309829/2022-4 awarded by the CNPq. W.S. acknowledges financial support from CNPq (Proc. 318052/2021-0, 300343/2022-1, 300834/2023-3, 301366/2023-3, 300252/2024-2 and 301472/2024-6).

\bibliography{bibliography}

\appendix

\section{S4GUI software description} \label{sec:appendixA}

In this section, we describe the class diagram and functional block diagram of S4GUI. These are two products generated during the planning stage of the software and were developed to assist during its development.

\subsection{Classes diagram} \label{subsec:S4GUI_class_diagram}

\begin{figure*}[tbh]
    \centering
    \includegraphics[width=\textwidth]{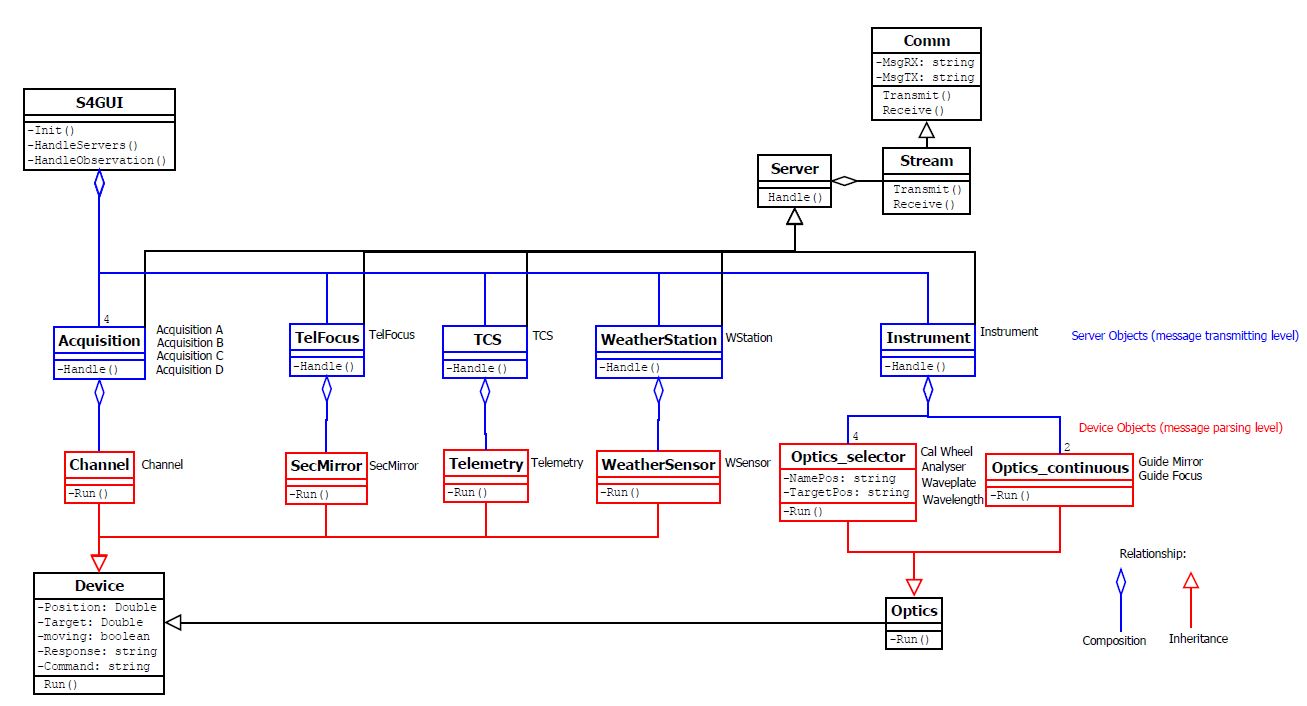}
    \caption{S4GUI class diagram. In the figure, the server (blue) and device (red) main group of classes are highlighted. The server classes inherit from the abstract class named ``Server" and implement communication routines. The device classes inherit from the abstract class ``Device" and represent the devices with which the S4GUI communicates. Both these group of classes have a composition relationship with S4GUI.}
    \label{fig:guiclasses}
\end{figure*}

Figure \ref{fig:guiclasses} illustrates the S4GUI class diagram, highlighting two primary groups: the server classes (blue) and the device classes (red). The device classes inherit from the abstract class named ``Device" and they were designed to represent the applications with which S4GUI interfaces. The Telemetry and SecMirros classes represent the telescope and its focuser system, respectively. The WeatherSensor class represents the weather station. The Optics\_selector and Optics\_continuous classes represent the polarimetric and guiding modules, respectively. The Channel class represents an instrument channel. These last three classes, in particular, have specific routines for the control of the instrument subsystem they represent. 

The server classes inherit from the abstract class named ``Server", which implements the communication counterpart of the Device classes. This communication is implemented using a data structure named queue, and two queues are created for each server instance: one for transmission and another for receiving the data. In particular, four instances of the Acquisition class are used to communicate with the four S4ACS instances.
Together, the Device and Server classes make up a two-layer abstraction used to represent the devices managed by S4GUI, contributing to the flexibility and expandability of the code.

Running in parallel to S4GUI, there is a standalone module that is used to interface the communication between this software and the external applications. Communication with S4GUI is implemented using the same queue data structure of the Server classes. Similarly, communication with external applications is implemented through the Ethernet network, using the ZeroMQ LabVIEW library. This configuration is used to separate the processes related to the communication from the control of the instrument, allowing S4GUI to continue to run in case of any failure that might occur when communicating with other applications.

\subsection{Functional block diagram} \label{subsec:S4GUI_functional_diagram}

\begin{figure*}[tbh]
    \centering
    \includegraphics[scale=0.75]{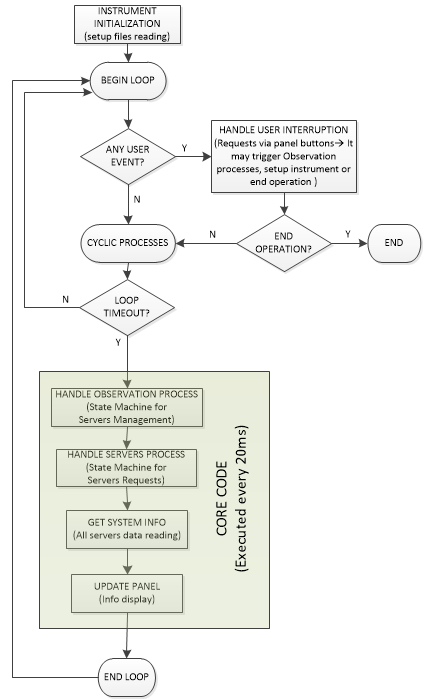}
    \caption{S4GUI functional block diagram.}
    \label{fig:guidiagram}
\end{figure*}

Figure \ref{fig:guidiagram} shows the functional block diagram executed by S4GUI during its operation. This diagram can be divided into two main parts based on the tasks performed. The first part is an event loop structure that handles user interactions such as pressing the start button to trigger the acquisition of a series of images. The second part, enclosed within the ``CORE CODE" rectangle, operates at regular 20~ms intervals and contains the main routines related to the instrument control. This part is composed of four functional blocks: {\sc HANDLE OBSERVATION PROCESS}, {\sc HANDLE SERVERS PROCESS}, {\sc GET SYSTEM INFO}, and {\sc UPDATE PANEL}. The {\sc HANDLE OBSERVATION PROCESS} is responsible for communicating with the applications that S4GUI interfaces and for handling any inconsistency that might occur during the acquisition of a series of images. The {\sc HANDLE SERVERS PROCESS} block executes the requests generated in the {\sc HANDLE OBSERVATION PROCESS} block and handles communication procedures. The {\sc GET SYSTEM INFO} block collects status information retrieved by these applications, while the {\sc UPDATE PANEL} block writes them to the S4GUI front panel indicators.

\end{document}